\documentstyle[twocolumn,aps]{revtex}

\begin{document}

\twocolumn[\hsize\textwidth\columnwidth\hsize\csname
               @twocolumnfalse\endcsname

\title{Gauge theory of a massive relativistic spinning point particle}

\author{J. H. Lorentsen$^\dagger$
and
N. K. Nielsen$^\ast$\\
Fysisk Institute, Odense University, DK-5230 Odense M, Denmark}

\maketitle

\begin{abstract}
  A massive relativistic spinning point particle in any number of
  dimensions has in a previous article been shown to
  be described by first class constraints, which define a gauge
  theory.  In the present paper we find the corresponding finite gauge
  transformations.
  By comparing the integrated gauge transformations to transformation
  equations found by Pryce, we conclude that the
  selection of gauge corresponds to selection of the relativistic
  center of mass frame in the model of Pryce, where a spinning
  particle is considered a composite object.  The Lorentz group is identified
  as the gauge group,
  and as gauge field we identify the relativistic angular velocity.
  We also show that an
  analogous physical interpretation is possible for the relativistic
  spherical top of Hanson and Regge.
\end{abstract}

\vskip2pc]

%%%%%%%%%%%%%%%%
\section{Introduction}
\label{intro}
%%%%%%%%%%%%%%%%

The problem of defining a
relativistic center of mass coordinate frame was considered by
Pryce \cite{pryce} many years ago. Demanding that
the definition of the center of mass in the non-relativistic limit
should reduce to the non-relativistic result, he found three different
ways in which the center of mass could be defined.  The view adapted
by Pryce is that a massive spinning particle is composed
of several objects. Spin is generated by the relative motion of the
objects, and their center of mass defines the position of the spinning
particle. Thus the orbital angular momentum, and hence also the spin,
depends on the choice of center of mass, since the total angular
momentum has to be independent of this choice. However,
the center of mass is not a unique concept in special relativity.

To describe a particle with spin formally, one can attach a rotating
frame to the particle \cite{itz:voros}. This corresponds to using the
Lorentz transformation matrix
$\Lambda^\mu\raisebox{-0.35ex}{\scriptsize$\nu$}$ as a dynamical
variable on the same footing as the particle position $x^\mu$
\cite{han:reg}. In \cite{ulrik:niels} it was shown to be convenient to
use canonical coordinates on the Poincar\'e group manifold as
dynamical variables instead of
$\Lambda^\mu\raisebox{-0.35ex}{\scriptsize$\nu$}$ and $x^\mu$ $^\bullet $

Pryce determined three ways in which the choice of center of
mass reference frame can be expressed through a set of constraints on
the spinning particle $^\ddagger$.  The so-called Pryce
constraint is
\begin{eqnarray}
\hat{S}^{\mu\nu}P_\nu =0, \label{eq:pryce-constr}
\end{eqnarray}
where $\hat{S}^{\mu\nu}$ is the spin matrix, and $P^\mu$ are the momenta.
This corresponds to a center of mass computed in the rest frame of
the composite body and next Lorentz transformed to an arbitrary frame.

 \vspace{1 mm}

\begin{center}
\line(1,0){200}
\end{center}

\vspace{1 mm}

\noindent$ ^\dagger$Electronic address: jhl@dirac.fys.ou.dk\\
$^\ast$Electronic address: nkn@fysik.ou.dk\\
$^\bullet$Our metric is $\eta^{\mu\nu}=\mbox{diag}(-1,1,\ldots,1)$,
  and we work in a $D$-dimensional space-time.\\
$^\ddagger$See \cite{han:reg} App. B for a useful
  review of the center of mass definitions.

\newpage

Hanson and Regge \cite{han:reg} Dirac quantized the point particle
using the Pryce constraint and the gauge condition (in the following
referred to as the Pryce gauge condition)
\begin{eqnarray}
  \Lambda^{0\mu}+\hbox{${\frac{P^\mu}{m}}$}=0 \label{prycegauge}
\end{eqnarray}
where $m$ is the mass. The resulting Dirac brackets between
coordinates $\hat{x}^\mu$, momenta $P^\mu$, and spin
$\hat{S}^{\mu\nu}$ are the same as the Poisson brackets Pryce found
for the corresponding quantities in his spinning particle model.  The
constraints (\ref{eq:pryce-constr}) are second class and hence do not
define a gauge theory.

The second set of spin constraints found by Pryce are
\begin{eqnarray}
\d S^{0\mu}=0,\label{eq:S0mu-constr}
\end{eqnarray}
where $\d S^{\mu\nu}$ is the corresponding spin matrix.  These
constraints correspond to a center of energy computed in the reference
frame of an arbitrary observer.  We will in the following refer to
these constraints as the $\d S^{0\mu}$-constraints.

The third  set of spin the constraints found by Pryce  are
\begin{eqnarray}
  (P^0+m)S^{0\mu}+S^{\mu i}P_i =0,\label{eq:wigner-constraint}
\end{eqnarray}
(referred to in the following as the Wigner constraints) where
$S^{\mu\nu}$ is the corresponding spin matrix. These constraints arise
when the center of mass is a weighted averages of the previous two
centers of mass; they
were also found by Wigner when  determining the representations of
the Poincar\'e group \cite{wigner}. If one uses Wigner's construction
of the representations of the Poincar\'e group by means of little
groups, and takes the one corresponding to the rest frame, an
identification of the spin matrix reveals the constraints
(\ref{eq:wigner-constraint}).

Modified constraints that include all three cases were found in
\cite{ulrik:niels}. The constraints are
\begin{eqnarray}
  \psi^\mu = S^{\mu\nu}(P_\nu
  -m\Lambda^0\raisebox{-0.35ex}{\scriptsize$\nu$}).
\label{eq:constraints}
\end{eqnarray}
These constraints were found to be first class constraints, and thus
define a gauge theory. Gauges should be selected by fixing $\Lambda^{0\mu}$.
The constraints
(\ref{eq:constraints}) are thus gauge dependent.  If we select the
gauge (\ref{prycegauge}), eq.~(\ref{eq:constraints}) reduces to the
Pryce condition (\ref{eq:pryce-constr}). In the same manner the Wigner
condition turns up if we select the gauge condition (referred to in
the following as the Wigner gauge condition)
\begin{eqnarray}
  \Lambda^{0\mu}- \eta^{0\mu}=0,
\end{eqnarray}
and for the $\d S^{0\mu}$-condition (\ref{eq:S0mu-constr}) we correspondingly
select the gauge condition (the $\d S^{0\mu}$-gauge condition)
\begin{eqnarray}
  \Lambda^{0\mu}-\hbox{$\frac{2P^0}{m}$}\eta^{0\mu}-\hbox{$\frac{P^\mu}{m}$}=0.
\end{eqnarray}
We should note here that this is a corrected version of
\cite{ulrik:niels} eq.(68). The above condition is chosen such that
the normalization
$\Lambda^{0\mu}\Lambda^0\raisebox{-0.35ex}{\scriptsize$\mu$} = -1$ is
valid.  The constraints are reducible since $(P_\lambda
-m\Lambda^0\raisebox{-0.35ex}{\scriptsize$\lambda$})\psi^\lambda=0$,
and thus the number of free constraints is $D-1$, and not $D$.

The purpose of this article is first to develop the gauge theory arising
from the result of \cite{ulrik:niels} that an arbitrary massive
spinning particle can be described by the (gauge-dependent) first-class
constraints (\ref{eq:constraints}).
The gauge field is found to be the
(generalized) angular velocity, and the gauge group is determined as
the Lorentz group
$SO(D-1,1)$.

Secondly, we deal with the issue of finite gauge transformations.
This is more complicated here than for ordinary gauge theories
because the constraints are gauge dependent.
As a consequence of this, construction of finite gauge transformations
involves integration of nonlinear
differential equations and is hence more complicated than the analogous
construction for ordinary gauge theories.

Pryce found transformation equations between the different sets of
variables \footnote{Our coordinates and spin variables $\hat {x}^{\mu }$ and
  $\hat{S}^{\mu \nu }$ in the case where
  (\ref{eq:pryce-constr}) holds correspond to Pryce's coordinates
  $X^\mu$ and spin matrix $\Sigma^{\mu\nu}$ \cite{pryce}. In \cite{pryce}
  the Wigner- and $\d S^{0\mu}$-gauge variables $x^\mu,
  S^{\mu\nu}$ and $\d x^\mu, \d S^{\mu\nu}$ are denoted by
  $\tilde{q}^\mu, \tilde{S}^{\mu\nu}$ and $q^\mu, S^{\mu\nu}$
  respectively.}. In our notation they are given by
\begin{eqnarray}
  \hat{x} = \d x^i -\frac{1}{m^2}S^{ij} P^j \label{eq:pryce-xtrans}
\end{eqnarray}
and
\begin{eqnarray}
  \hat{x}^i=x^i-
  \frac{\hat{S}^{ik}P_k}{P^0(P^0+m)}.\label{eq:pryce-Strans}
\end{eqnarray}
For the spin matrices the transformation equations are
\begin{eqnarray}
  \hat{S}^{kl}&=&\d S^{kl}+ \frac{P^m}{m^2}(\d S^{km}P^l-\d S^{lm}P^k)
\label{eq:pryce-Sd-Sh}
\end{eqnarray}
and
\begin{eqnarray}
S^{kl}+\frac{P_m}{P^0
  (P^0+m)}(\hat{S}^{km}P^l-\hat{S}^{lm}P^k)\label{eq:pryce-S-Sh}
\end{eqnarray}
We show below that the finite gauge transformation equations one gets
from the general constraints (\ref{eq:constraints}) are the
transformation equations
(\ref{eq:pryce-xtrans})--(\ref{eq:pryce-S-Sh}) that Pryce found for
the transformation between the different sets of variables. Thus the
selection of gauges in the gauge theory defined by the constraints
(\ref{eq:constraints}) corresponds to a definition of the
relativistic center of mass frame.

The relativistic spherical top defined in \cite{han:reg} is also
considered, and a similar construction is carried out for this system.

The layout of the paper is the following: In sec. II we review the Poisson
 brackets found in \cite{han:reg} and \cite{ulrik:niels}, with emphasis on
the first class constraint algebra found in \cite{ulrik:niels}. In sec. III
we determine the infinitesimal gauge transformations generated by the
constraints.
In sec. IV the infinitesimal gaugetransformations are integrated and
the connection to the transformation formulas of Pryce established.
Finally the analysis is extended to the relativistic spherical top in sec. V.
Concluding remarks are contained in sec. VI.
%%%%%%%%%%%%%%%%%%%%%%%
\section{Poisson brackets}
\label{poiss}
%%%%%%%%%%%%%%%%%%%%%%%

The determination of whether a set of constraints is first or second
class begins by the determination of the Poisson brackets involving the
variables in the theory. In \cite{ulrik:niels} the relevant Poisson brackets
 were found from  the Poincar\'e group structure equations and the
canonical commutation relations for
canonical coordinates on the group manifold and their momenta:
\begin{eqnarray}
  \{S_{\mu\nu},S_{\lambda\rho}\}&=& \hbox{$\frac{1}{2}$}
  C^{\kappa\delta}\hspace{0.1 mm}_{\mu\nu,\lambda\rho}
S_{\kappa\delta}\nonumber\\
  \{S_{\mu\nu},\Lambda^{\lambda}\raisebox{-0.35ex}{\scriptsize$\rho$}
  \}&=& C^\kappa \hspace{0.1 mm}_{\mu\nu,\rho}
  \Lambda^\lambda\raisebox{-0.35ex}{\scriptsize$\kappa$}
  \label{eq:brackets1}
\end{eqnarray}
where $S^{\mu\nu}$ denotes the spin matrix in generality.
 The rest of the brackets vanish
\begin{eqnarray}
  \{\Lambda^{\mu\nu},\Lambda^{\lambda\rho}\} =
  \{\Lambda^{\mu\nu},P^\rho\} = \{\Lambda^{\mu\nu},x^\lambda\} =
  0,\nonumber\\ \{ S^{\mu\nu}, P^\lambda\} = \{ S^{\mu\nu},
  x^\lambda\} = 0,\nonumber\\ \{x^\mu,x^\nu\} = 0.\label{eq:brackets2}
\end{eqnarray}
Here the Poincar\'e group
structure constants are
\begin{eqnarray}
  C^{\mu\nu}\hspace{0.1 mm}_{\lambda\rho,\kappa\delta}&=&
  \delta^\mu\raisebox{-0.35ex}{\scriptsize$\kappa$}
  C^{\nu}\hspace{0.1 mm}_{\lambda\rho,\delta}
  -\delta^\nu\raisebox{-0.35ex}{\scriptsize$\kappa$}
  C^{\mu}\hspace{0.1 mm}_{\lambda\rho, \delta}
   \nonumber \\
   &-&
  \delta^{\mu}\raisebox{-0.35ex}{\scriptsize$\delta$}
  C^{\nu}\hspace{0.1 mm}_{\lambda\rho,\kappa} +
  \delta^\nu\raisebox{-0.35ex}{\scriptsize$\delta$}
  C^{\mu}\hspace{0.1 mm}_{\lambda\rho,\kappa},
  \nonumber\\
  C^\lambda \hspace{0.1 mm}_{\mu\nu,\rho}&=&
  \delta^\lambda\raisebox{-0.35ex}{\scriptsize$\nu$} \eta_{\mu\rho}-
  \delta^\lambda\raisebox{-0.35ex}{\scriptsize$\mu$} \eta_{\nu\rho}.
\end{eqnarray}
The constraints are
\begin{eqnarray}
  \phi=P^2 +m^2, \hspace{1cm} \psi^\mu=S^{\mu\lambda}(P_\lambda
  -m\Lambda^0\raisebox{-0.35ex}{\scriptsize$\lambda$}).
   \label{CON}
\end{eqnarray}
$\phi$ is the mass-shell condition. This is a system of first class
constraints since $\{\phi,\phi\}=\{\phi,\psi^\mu\}=0$ and
\begin{eqnarray}
  \{\psi^\mu,\psi^\nu\}= P^\mu\psi^\nu-P^\nu\psi^\mu+S^{\mu\nu}\phi.
\end{eqnarray}
The constraint algebra is closed (the constraints are first class)
and therefore they define a gauge theory.

The relativistic spherical top of Hanson and Regge \cite{han:reg} is defined by
a system of
constraints where the mass-shell constraint is replaced by the condition
\begin{eqnarray}
  \phi=P^2+f(\hbox{$\frac{1}{2}$} S_{\mu\nu}S^{\mu\nu})=0
   \label{Regge}
\end{eqnarray}
which incorporates Regge trajectories. Here the function
$f$ is real but otherwise unspecified. In \cite{ulrik:niels} the mass-shell
condition was
shown to have the general form
\begin{eqnarray}
  \phi &=& P^2+ f(m^2), \hspace{3mm} m^2 = \hbox{$\frac{1}{2}$}
S_{\mu\nu}S^{\mu\nu}+S^{\mu\lambda}\Lambda^0\raisebox{-0.35ex}{\scriptsize$\
lambda$}
S_\mu\raisebox{0.95ex}{\scriptsize$\rho$}\Lambda^0\raisebox{-0.35ex}
{\scriptsize$\rho$},
  \nonumber\\ \psi^\mu &=&
  S^{\mu\nu}(P_\nu-m\Lambda^0\raisebox{-0.35ex}{\scriptsize$\nu$})
  \label{eq:sph-constraints}.
\end{eqnarray}
which by the Pryce gauge condition reduces to (\ref{Regge}).
If we use the Poisson brackets (\ref{eq:brackets1}) and
(\ref{eq:brackets2}) we find
\begin{eqnarray}
  \{m,x^\mu\}&=& 0\nonumber\\ \{m,S^{\mu\nu}\}&=& 0\nonumber\\
  \{m,\Lambda^{\mu\nu}\}&=& \hbox{$\frac{1}{m}$}\big(
  \Lambda^{\mu\lambda}S^\nu\raisebox{-0.35ex}{\scriptsize$\lambda$}\nonumber\\
  & &\ \
  +\eta^{0\mu}S^\nu\raisebox{-0.35ex}{\scriptsize$\rho$}\Lambda^{0\rho}
  -\Lambda^{\mu\kappa}\Lambda^{0\nu}
  S_{\kappa\rho}\Lambda^{0\rho}\big).\label{eq:brackets3}
\end{eqnarray}
From the above equations we have $\{m,\Lambda^{0\mu}\}=0$. This
means that the Poisson brackets $\{ \phi,\phi\}$, $\{\phi,\psi^\mu\}$,
and $\{ \psi^\mu,\psi^\nu\}$ are the same as the corresponing ones in
the point particle case. The constraints are thus still first class and
still define a gauge theory. The main difference to the
point particle case is that the mass is now a function of the spin.

%%%%%%%%%%%%%%%%%%%%%%%
\section{Infinitesimal gauge transformations}
%%%%%%%%%%%%%%%%%%%%%%%

In this section we determine the infinitesimal
gauge transformations generated by the system of constraints
(\ref{CON}), in order to find the gauge transformation
equations. We first consider the case of a spinless particle where we
only have the mass shell constraint.  This simple and well-known
example serves to display the general strategy that will be used for
the spinning particle.

%%%%%%%%%%%%%%%%%%%%%%%%%%%%%%
\subsection{The spinless point particle}
%%%%%%%%%%%%%%%%%%%%%%%%%%%%%%

The spinless point particle is described by the coordinates $x^\mu$
and momenta $P^\mu$. The only constraint is the mass shell condition
$\phi$ in eq.~(\ref{CON}).  Since $\{\phi,\phi\}=0$ this is a
first class system.  Time development is found from the Hamiltonian
\begin{eqnarray}
  H=\hbox{$\frac{e}{2}$}(P^2+m^2)
\end{eqnarray}
where $e$ is a Lagrange multiplier, such that for an arbitrary
dynamical variable $X$ the time derivative is
\begin{equation}
  \dot{X}=\{ X,H\}.
\end{equation}
In a similar way we find the gauge transformations $\delta X$ is found
from the generator $Q$:
\begin{eqnarray}
  Q=\hbox{$\frac{\alpha}{2}$}(P^2+m^2)
\end{eqnarray}
(with $\alpha $ infinitesimal and real) by the formula
\begin{equation}
  \delta X=\{X,Q\}.
\end{equation}
Using these equations we find that the equations of motion are
\begin{eqnarray}
  \dot{x}^\mu = eP^\mu, \hspace{.5cm} \dot{P}^\mu =
  0, \hspace{.5cm} \dot{e}=0, \label{eq:mellem9}
\end{eqnarray}
and the gauge transformations are
\begin{eqnarray}
  \delta x^\mu = \alpha P^\mu, \hspace{.5cm} \delta P^\mu = 0,
\hspace{.5cm} \delta e=0.
\label{eq:xp-trans}
\end{eqnarray}
From $\delta (\dot{x}^\mu)= \dot{(\delta x^\mu)}$ follows immediately
\begin{eqnarray}
  \delta e = \dot{\alpha},
\end{eqnarray}
which is the gauge transformation of an Abelian gauge field in one
dimension.  Thus we have an Abelian gauge theory, where $e$ is the
gauge field. It is a constant, and $\alpha $ is consequently linear in $\tau $.

We can immediately integrate the infinitesimal gauge transformations
to obtain finite gauge transformations. As a specific example we can
take $\alpha=0$ in the covariant gauge where
$x=x_{\hbox{\scriptsize{cov}}}$ with $\dot
x^2_{\hbox{\scriptsize{cov}}}+1=0$, and thus $e=\frac{1}{m}$ from
(\ref{eq:mellem9}) and the mass-shell condition. Equation
(\ref{eq:xp-trans}) can be integrated to
\begin{eqnarray}
  x^\mu(\alpha) = x^\mu_{\hbox{\scriptsize{cov}}}+\alpha
  P^\mu.\label{eq:cov.gau.tr}
\end{eqnarray}
The selection of the gauge determines the Lagrange multiplier $e$.
Taking the time derivative of (\ref{eq:cov.gau.tr}), we get
\begin{eqnarray}
  eP^\mu &=& \dot{x}^\mu_{\hbox{\scriptsize{cov}}}+\dot{\alpha}P^\mu.
\end{eqnarray}
The equation of motion of $x^\mu_{\hbox{\scriptsize{cov}}}$ then gives
us that
\begin{eqnarray}  e=\hbox{$\frac{1}{m}$}+\dot{\alpha}.\end{eqnarray}
As an example we could choose the proper time gauge $x^0=\tau$.  The
Lagrange multiplier and gauge parameter are then by (\ref{eq:mellem9})
\begin{eqnarray}
  e=\hbox{$\frac{1}{P^0}$}, \hspace{1.5cm}\dot{\alpha}=
  \hbox{$\frac{1}{P^0}$}-\hbox{$\frac{1}{m}$}.
\end{eqnarray}
We found above that the constraint gives an Abelian gauge
theory. We then determined the general transformation formula
(\ref{eq:cov.gau.tr}) allowing a transformation  from the covariant gauge to
any other gauge. \\

%%%%%%%%%%%%%%%%%%%%%%%%%%%%%%
\subsection{A point particle with spin\label{sec33}}
%%%%%%%%%%%%%%%%%%%%%%%%%%%%%%

For a particle with spin, we generate infinitesimal gauge
transformations from $\delta X=\{X,Q\}$ where $X$ is an arbitary
variable, and $Q$ is a linear combination of the
constraints in eq. (\ref{CON}):
\begin{eqnarray}
  Q=\hbox{$\frac{\alpha}{2}$}(P^2+m^2)+ \alpha^\mu
  S_{\mu \nu }(P^\nu-m\Lambda^{0\nu})
\end{eqnarray}
where $\alpha$ and $\alpha^\mu$ are arbitary infinitesimal functions.
Using the Poisson brackets in (\ref{eq:brackets1}) and
(\ref{eq:brackets2}) gives us the infinitesimal gauge transformations:
\begin{eqnarray}
    \delta x^\mu&=& \alpha P^\mu + \alpha^\lambda
     S_\lambda\raisebox{0.95ex}{\scriptsize$\mu$}\label{eq:xgauge}\\
    \delta S^{\mu\nu}&=& \alpha^\mu \psi^\nu -\alpha^\nu \psi^\mu
     +\alpha^\lambda S^\mu\raisebox{-0.35ex}{\scriptsize$\lambda$} P^\nu
     -\alpha^\lambda S^\nu\raisebox{-0.35ex}{\scriptsize$\lambda$}
     P^\mu
     \label{eq:Sgauge}\\
    \delta \Lambda^{\mu\nu}&=& \Lambda^\mu\raisebox{-0.35ex}{\scriptsize$\rho$}
     \lambda(\alpha)^{\rho\nu}\label{eq:lambdatrans}\\
    \delta P^\mu &=& 0, \hspace{0.5 mm}
    \delta M^{\mu\nu} = 0\label{eq:inf-trans}
\end{eqnarray}
where we defined
\begin{eqnarray}
  \lambda(\alpha)^{\mu\nu}=\alpha^\mu(P^\nu-m\Lambda^{0\nu})-
  \alpha^\nu(P^\mu-m\Lambda^{0\mu}).
  \label{lambdaalpha}
\end{eqnarray}
We notice that the generators of the Poincar\'e group are
invariant. We also notice that all the equations, except
eq.~(\ref{eq:lambdatrans}), are linear differential equations.
(\ref{eq:lambdatrans}) is nonlinear bacause the constraints
(\ref{eq:constraints}) are gauge dependent.

Since the constraints are first class,
they are invariant under gauge transformations,
after the constraints are imposed.

The
Hamiltonian is a linear combination of the constraints
\begin{eqnarray}
  H=\hbox{$\frac{e}{2}$}(P^2+m^2)+v^\mu
  S_{\mu\nu}(P^\nu-m\Lambda^{0\nu}).
\end{eqnarray}
We determine the time derivative of any parameter $X$ by means of
$\dot{X}=\{X,H\}$. The outcome is (cf. (\ref{eq:xgauge})-(\ref{eq:inf-trans})):
\begin{eqnarray}
     \dot x^\mu&=& e P^\mu + v^\lambda
     S_\lambda\raisebox{0.95ex}{\scriptsize$\mu$}\label{eq:xdot}\\
     \dot S^{\mu\nu}&=& v^\mu \psi^\nu -v^\nu \psi^\mu
     +v^\lambda S^\mu\raisebox{-0.35ex}{\scriptsize$\lambda$} P^\nu
     -v^\lambda S^\nu\raisebox{-0.35ex}{\scriptsize$\lambda$}
     P^\mu
     \label{eq:Sdot}\\
     \dot \Lambda^{\mu\nu}&=& \Lambda^\mu\raisebox{-0.35ex}{\scriptsize$\rho$}
     \lambda(v)^{\rho\nu}\label{eq:lambdadot}\\
    \dot v^{\mu }&=&0, \label{vdot} \\
    \dot P^\mu &=& 0, \hspace{0.5 mm}
    \dot M^{\mu\nu} = 0.\label{eq:inf-dot}
\end{eqnarray}
where
\begin{eqnarray}
  \lambda(v)^{\mu\nu}=v^\mu(P^\nu-m\Lambda^{0\nu})-
  v^\nu(P^\mu-m\Lambda^{0\mu}). \label{eq:lambdav}
\end{eqnarray}
$\lambda(v)^{\mu\nu}$ was in \cite{han:reg} identified as a relativistic
analog of the angular momentum.
By means of the Jacobi identity and the constraints we find that
carrying out a gauge transformation on $\dot{\Lambda }^{\mu\nu}$ is the
same as taking time derivative of $\delta\Lambda^{\mu\nu}$.
Using (\ref{eq:lambdatrans}) and (\ref{eq:lambdadot}), we find
\begin{eqnarray}
  \delta\lambda(v)^{\mu\nu} &=&
  \dot{\lambda}(\alpha)^{\mu\nu}+
  \hbox{$\frac{1}{2}$}\hbox{$\frac{1}{2}$}
  C^{\mu\nu}\hspace{0.1 mm}_{\lambda\rho,\sigma\tau} \lambda(v)^{\lambda\rho}
  \lambda(\alpha)^{\sigma\tau}.
  \label{eq:gauge.field}
\end{eqnarray}
This is a transformation equation for a Yang--Mills field in one dimension,
where the Lorentz group
$SO(D-1,1)$ is the gauge group.

Eq.~(\ref{eq:gauge.field}) is equivalent to the following gauge transformation
of the Lagrange multipliers:
\begin{eqnarray}
  \delta v^\mu=\dot\alpha^\mu+v^\mu(\alpha\cdot P)-\alpha^\mu(v\cdot
  P).
\label{eq:vtrns}
\end{eqnarray}
Using (\ref{vdot}), we conclude from (\ref{eq:vtrns}) that the gauge parameters
$\alpha ^\mu $ have two  options:
\begin{itemize}
\item They can be linear in $\tau $ and proportional to
$v^{\mu }$. Gauge transformations in this class have the same effect on
the variables as the time development
described by (\ref{eq:xdot}), (\ref{eq:Sdot}), (\ref{eq:lambdadot} and
(\ref{eq:inf-dot}), and can thus also be used to eliminate $v^{\mu }$ in
these equations.
\item  They can be constant
(independent of $\tau $). These global gauge gauge transformations are
integrated in
the following section to obtain generalizations of the transformation formulas
mentioned in the introduction (\ref{eq:pryce-xtrans}), (\ref{eq:pryce-Strans}),
(\ref{eq:pryce-Sd-Sh}) and (\ref{eq:pryce-S-Sh})).
\end{itemize}

%%%%%%%%%%%%%%%%%%%%%%
\section{Finite gauge transformations}
%%%%%%%%%%%%%%%%%%%%%%

In this section we consider integration of the nontrivial infinitesimal gauge
transformation formulas found in sec. IIIB to obtain the corresponding finite
gauge transformations.

We assume  $\dot{\alpha}^\mu=0$ (global gauge transformations).

%%%%%%%%%%%%%%%%%%%%%%
\subsection{Finite gauge transformations of spin matrix, coordinates,
  and Lagrange multipliers}
%%%%%%%%%%%%%%%%%%%%%%

In order to find the finite transformations for the spin matrix, we use
eq.~(\ref{eq:Sgauge})  with constraints taken
into account ($\psi ^{\mu }=0$)
\begin{eqnarray}
  \delta S^{\mu\nu} =\alpha^\lambda
  S^\mu\raisebox{-0.35ex}{\scriptsize$\lambda$} P^\nu -\alpha^\lambda
  S^\nu\raisebox{-0.35ex}{\scriptsize$\lambda$} P^\mu
  \nonumber.
\end{eqnarray}
We write the infinitesimal variables
$\alpha^\mu$ as follows
\begin{eqnarray}
  \alpha^\mu =\delta s\;a^\mu,
\end{eqnarray}
where $s$ is a real ``gauge development'' parameter, and $a^\mu$ is
independent of $s$. This gives
\begin{eqnarray}
  \frac{d S^{\mu\nu}(s)}{ds}=a^\lambda
  S^\mu\raisebox{-0.35ex}{\scriptsize$\lambda$}(s) P^\nu - a^\lambda
  S^\nu
  \raisebox{-0.35ex}{\scriptsize$\lambda$}P^\mu.\label{eq:mellem20}
\end{eqnarray}
whence
\begin{eqnarray}
  \frac{d(S^{\mu\rho}(s)a_\rho)}{ds}=a_\rho P^\rho S^{\mu\lambda}(s)
  a_\lambda.
\end{eqnarray}
that is integrated
\begin{eqnarray}
  S^{\mu\lambda}(s)a_\lambda = S^{\mu\lambda}(0)a_\lambda e^{a\cdot
    P\, s}.\label{eq:mellem42}
\end{eqnarray}
Insertion back into eq.~(\ref{eq:mellem20}), which next is integrated, gives
\begin{eqnarray}
S^{\mu\nu}(s)&-&S^{\mu\nu}(0) \nonumber \\
&=&\frac{a_\lambda}{a\cdot
  P}(S^{\mu\lambda}(0)P^\nu -S^{\nu\lambda}(0)P^\mu)(e^{a\cdot P
s}-1).\label{eq:mellem31a}
\end{eqnarray}
This equation has already the same form as (\ref{eq:pryce-Sd-Sh}) and
(\ref{eq:pryce-S-Sh}). It takes a simpler form in the
limit $s\rightarrow \infty$,
if we select the sign of $a^\mu$ such that $a\cdot P<0$:
\begin{eqnarray}
  S^{\mu\nu}(\infty)=S^{\mu\nu}(0)-\frac{a_\lambda}{a\cdot P}(
  S^{\mu\lambda}(0)P^\nu - S^{\nu\lambda}(0)P^\mu)\label{eq:mellem34}
\end{eqnarray}
From this equation we see that $S^{\mu\nu}(\infty)$ fulfil the
constraints $ S^{\mu\lambda}(\infty)a_\lambda=0$. If we compare these
constraints with the constraints (\ref{eq:constraints}) we conclude
\begin{eqnarray}
   P^\mu -m\Lambda^{0\mu}(\infty)\label{eq:align}=Ca^{\mu }
   \label{C}
\end{eqnarray}
where $C$ is a real proportionality factor that is determined in terms of
$a^{\mu }$ and $P^{\mu }$ in sec. IV.C.

For the coordinates we proceed as we did for the spin matrix.  If we
use $\alpha^\mu=\delta s\; a^\mu$ and $\alpha =0$ in (\ref{eq:xgauge})
and insert the result we found for the spin matrix $S^{\mu\nu}(s)$, we
get
\begin{eqnarray}
  \frac{d x^\mu(s)}{ds} = a^\rho
  S_\rho\raisebox{0.95ex}{\scriptsize$\mu$}(0) e^{a\cdot P s}.
  \label{eq:dxdt}
\end{eqnarray}
that is integrated
\begin{eqnarray}
  x^\mu(s)=\frac{a^\rho}{a\cdot P}
  S_\rho\raisebox{0.95ex}{\scriptsize$\mu$}(0)(e^{a\cdot P s} -1)+
  x^\mu(0).
\end{eqnarray}
If we again use that $a\cdot P<0$, and take the $s\rightarrow \infty$
limit, we get the transformation equations
\begin{eqnarray}
  x^\mu(\infty)=x^\mu(0)+\frac{a^\rho}{a\cdot
    P}S^\mu\raisebox{-0.35ex}{\scriptsize$\rho$}(0).
\label{eq:mellem35}
\end{eqnarray}

We finally find the finite gauge transformations of the Lagrange
multipliers. From eq. (\ref{eq:gauge.field}) follows that the quantity
\begin{equation}
\zeta=v\cdot P
\label{defzeta}
\end{equation}
is invariant under global gauge transformations
Consequently (\ref{eq:gauge.field}) implies
\begin{eqnarray}
  \frac{d v^\mu(s)}{d s}=v^\mu(s)(a\cdot P)-\zeta a^\mu .
  \label{eq:mellem19a}
\end{eqnarray}
that is integrated
\begin{eqnarray}
  v^\mu(s)=\frac{\zeta  a^\mu }{a\cdot P}
  +(v^\mu(0)-\frac{\zeta  a^\mu }{ a\cdot P})e^{a\cdot P\,
    s} \label{eq:vint.pp}
\end{eqnarray}
that for $a\cdot P<0$ in the limit
$s\rightarrow\infty$ reduces to
\begin{eqnarray}
  v^\mu(\infty)=\frac{\zeta
    a^\mu}{a\cdot P}.
  \label{eq:mellem44}
\end{eqnarray}
The proportionality factor $\zeta $ is undetermined.

%%%%%%%%%%%%%%%%%%%%%%
\subsection{Connection to the Pryce transformation formulas}
%%%%%%%%%%%%%%%%%%%%%%

In the previous subsection a general formalism for finite gauge
transformations was constructed. We next demonstrate that the
transformation formulas of Pryce \cite{pryce} quoted in section
\ref{intro} come out as special cases.

We begin by considering the gauge transformations, where the Pryce
gauge (\ref{prycegauge}) corresponds to $s=\infty $. This means $a^\mu
= \alpha P^\mu$ , with $\alpha $ positive to ensure $a\cdot P<0$, but
otherwise arbitrary. From
(\ref{eq:mellem34}) and (\ref{eq:mellem35}) we get
\begin{eqnarray}
  \hat{S}^{\mu\nu}=S^{\mu\nu}(0)+\frac{P_\lambda}{m^2}(
  S^{\mu\lambda}(0)P^\nu - S^{\nu\lambda}(0)P^\mu)\label{eq:mellem36}
\end{eqnarray}
\begin{eqnarray}
  \hat{x}^{\mu}=x^\mu(0)-\frac{P_\lambda}{m^2}S^{\mu \lambda }(0).
  \label{eq:mellem37}
\end{eqnarray}
The variables $S^{\mu\nu}(0)$ and $x^{\mu }(0)$ have to be specified
in order to get Pryce's transformation formulas. If they are $\d
S^{\mu 0}$-gauge quantities, we get by taking $S^{\mu 0}(0)=0$:
\begin{eqnarray}
 \hat{S}^{\mu\nu}=\d S^{\mu\nu}
   +\frac{P_i}{m^2}(\d S^{\mu i}P^\nu - \d S^{\nu i}P^\mu).
  \label{eq:firstpryce1}
\end{eqnarray}
and
\begin{eqnarray}
 \hat{x}^\mu = \d x^\mu-\frac{P^i}{m^2} \d S^{\mu i} \label{eq:firstpryce2},
\end{eqnarray}
while by taking the initial variables to be Wigner-gauge quantities we
get by the Wigner constraint:
\begin{eqnarray}
  \hat{S}^{\mu\nu}=S^{\mu\nu}+\frac{P_i}{m(P^0+m)}( S^{\mu i}P^\nu -
  S^{\nu i}P^\mu)\label{eq:mellem38}
\end{eqnarray}
\begin{eqnarray}
  \hat{x}^{\mu}=x^\mu-\frac{P_i}{m(P^0+m)}S^{\mu i}.
\label{eq:mellem39}
\end{eqnarray}
where $i$ only runs over spatial indices.  These equations are
generalizations to arbitrary dimensionality of those Pryce
\cite{pryce} found for the corresponding center of mass reference
frames (cf. (\ref{eq:pryce-xtrans}) and (\ref{eq:pryce-Sd-Sh})).

Transformation equations, where the Wigner gauge corresponds to
$s=\infty $, can also be found. From the Wigner gauge conditions and
(\ref{eq:align}) one finds $a^\mu = \alpha (P^\mu-m\eta^{0\mu})$,
where again $\alpha>0$ to ensure $a\cdot P <0$.
The transformation equations (\ref{eq:mellem34}) and
(\ref{eq:mellem35}) in this case imply:
\begin{eqnarray}
  S^{\mu\nu}=S^{\mu\nu}(0)+\frac{(P_\lambda-m\eta _{0\lambda
      })}{m(P^0+m)}( S^{\mu\lambda}(0)P^\nu -
  S^{\nu\lambda}(0)P^\mu)\label{eq:mellem40}
\end{eqnarray}
\begin{eqnarray}
  x^\mu=x^\mu(0)-\frac{(P_\lambda-m\eta _{0\lambda })}{m(P^0+m)}S^{\mu
    \lambda }(0)..
\label{eq:mellem51}
\end{eqnarray}
Since we already have found the finite gauge transformation leading
from the Wigner gauge to the Pryce gauge, we only consider the case
where we start in the $\d S ^{\mu 0}$-gauge. Here we get:
\begin{eqnarray}
  S^{\mu\nu}=\d S^{\mu\nu}+\frac{P_i}{m(P^0+m)}( \d S^{\mu i}P^\nu -
  \d S^{\nu i}P^\mu)\label{eq:mellem43}
\end{eqnarray}
\begin{eqnarray}
  x^\mu=\d x^\mu-\frac{P_i}{m(P^0+m)}\d S^{\mu i}
\label{eq:mellem52}
\end{eqnarray}
which again are some of Pryce's transformation formulas.

%%%%%%%%%%%%%%%%%%%%%%%%%%%%
\subsection{Finite gauge transformation of the Lorentz transformations}
%%%%%%%%%%%%%%%%%%%%%%%%%%%%

Up to this point we have considered all variables with nontrivial
gauge transformation laws, except the Lorentz transformations
$\Lambda^{\mu\nu}$. As noted earlier we can not expect the
transformation equation for them to be as easy to integrate as the
other ones, because they involve nonlinear differential equations. On
the other hand, it is important to demonstrate that the finite gauge
transformations can be determined, because gauge fixing is done by
fixing $\Lambda ^{0\mu }$.  From eq.~(\ref{eq:lambdatrans}) with
$\alpha^\mu = \delta s\, a^\mu$ follows
\begin{eqnarray}
\frac{d\Lambda^{0\mu}}{ds}=\Lambda^0\raisebox{-0.35ex}{\scriptsize$\lambda$}
(a^\lambda
  P^\mu-a^\mu P^\lambda)
  -m(a^\mu+\Lambda^{0\mu}\Lambda^0\raisebox{-0.35ex}{\scriptsize$\lambda$}
  a^\lambda).
\label{eq:lambda-ref}
\end{eqnarray}
If we consider a boost in any direction we can write $\Lambda^{0\mu}$ as
\begin{eqnarray}
  \Lambda^{00}(s)=-\cosh{\psi(s)}&\hspace{8mm}&
  \Lambda^{0i}(s)=c^i(s)\sinh{\psi(s)}\nonumber\\ &\vec{c}(s)^2=1&
  \label{eq:lorentz.boost}
\end{eqnarray}
where $\psi$ and $\vec{c}$ are functions of $s$.
When this is inserted into eq.~(\ref{eq:lambdatrans}, we get
\begin{eqnarray}
  \frac{d\psi}{ds}&=&c^i(a^0P^i-a^iP^0)-ma^0\sinh
  \psi-m\cosh{\psi}c^ia^i
\label{eq:psitrans}
\end{eqnarray}
and
\begin{eqnarray}
  \frac{dc^i}{ds}\sinh{\psi}&=& (\delta ^{ij}-c^ic^j)(\cosh
  \psi(a^0P^j-a^jP^0)-ma^j)\nonumber \\ &+&(a^jP^i-a^iP^j)c^j \sinh
  \psi .
   \label{eq:citrans}
\end{eqnarray}
Eqs. (\ref{eq:psitrans}) and (\ref{eq:citrans}) are the final
general equations.

The solution of (\ref{eq:citrans}) is
\begin{eqnarray}
c^i=\varepsilon \frac{P^i}{|\vec{P}|}, \hspace{0.1 mm} \varepsilon =\pm 1
\label{eq:ditrans}
\end{eqnarray}
for
\begin{eqnarray}
a^{\mu }=\alpha P^{\mu }+\beta \eta ^{0\mu}
\label{eq:eitrans}
\end{eqnarray}
with $\alpha$ and $\beta$ arbitary. These gauge choices include
the three gauges mentioned in the introduction.

The general solution of (\ref{eq:psitrans}) in these gauges is found by
quadrature
\begin{eqnarray}
  e^{\psi(s)}= \frac{Y(s)
     (\frac{P^0-\varepsilon \mid\vec P\mid }{m})
    +(a^0-\varepsilon \frac{\vec{a}\cdot\vec{P}}{\mid \vec{P}\mid})
     (\frac{P^0+\varepsilon \mid\vec P\mid }{m})}
    {a^0+\varepsilon \frac{\vec{a}\cdot\vec{P}}{\mid \vec{P}\mid}-Y(s)}
\label{eq:psiintegral}
\end{eqnarray}
where
\begin{equation}
 Y(s)=\frac{(a^0+\varepsilon \frac{\vec{a}\cdot\vec{P}}
      {\mid \vec{P}\mid})e^{\psi (0)}
      -(a^0-\varepsilon \frac{\vec{a}\cdot\vec{P}}{\mid \vec{P}\mid})
     (\frac{P^0+\varepsilon \mid\vec P\mid }{m})} {
     e^{\psi (0)} +\frac{P^0-\varepsilon \mid\vec P\mid }{m}} e^{sa\cdot P}.
\end{equation}

We shall only give explicit expressions for $\Lambda ^{0\mu }(s)$ in the limit
$s\rightarrow \infty $. In this limt we have for $a\cdot P<0$ that
$Y(s)\rightarrow 0$
and consequently
\begin{eqnarray}
e^{\psi (\infty )}=
\frac{(a^0-\varepsilon \frac{\vec{a}\cdot\vec{P}}{\mid \vec{P}\mid})
     (\frac{P^0+\varepsilon \mid\vec P\mid }{m})}
    {a^0+\varepsilon \frac{\vec{a}\cdot\vec{P}}{\mid \vec{P}\mid}}.
\label{eq:Psiintegral}
\end{eqnarray}
Inserting this result into (\ref{eq:lorentz.boost}) and using
(\ref{eq:ditrans})-(\ref{eq:eitrans}) we obtain
\begin{eqnarray}
P^{\mu }-m\Lambda ^{0\mu }(\infty )=-2a^{\mu }\frac{a\cdot P}
{(a^0)^2-(\frac{\vec{a}\cdot \vec{P}}{\mid \vec{P} \mid })^2}
\end{eqnarray}
in conformity with (\ref{C}), with
\begin{eqnarray}
C=-\frac{2a\cdot P}
{(a^0)^2-(\frac{\vec{a}\cdot \vec{P}}{\mid \vec{P} \mid })^2}.
\end{eqnarray}

The remaining components of $\Lambda ^{\mu \nu }$ can be determined from
(\ref{eq:lambdatrans}) once the components of $\Lambda ^{0\mu }$ are known.
The resulting differential equations are linear.

%%%%%%%%%%%%%%%%%%%%
\section{The relativistic spherical top}
%%%%%%%%%%%%%%%%%%%%

We noticed in section \ref{poiss} that the relativistic spherical top
 is defined from a set of first class
constraints (\ref{eq:sph-constraints}). This means that the set of
constraints define a gauge
theory. We generate infinitesimal gauge transformations through
\begin{eqnarray}
  \delta X = \{X, Q\}\hspace{1.5cm} Q= \frac{\alpha }{2}\phi +\alpha_\mu
  \psi^\mu
\label{eq:sph-gauge-tr}
\end{eqnarray}
where $\phi$ and $\psi^\mu$ are defined in equations
(\ref{eq:sph-constraints}). Using (\ref{eq:brackets3}) one sees that
the infinitesimal gauge transformations and equations of motion of $x^\mu$,
$P^\mu$,
$M^{\mu\nu}$, and $S^{\mu\nu}$ are the same as the ones
(\ref{eq:inf-trans}) for the point particle.
The mass $m$ defined in (\ref{eq:sph-constraints}) is gauge invariant.

$m$ has non-vanishing Poisson brackets with $\Lambda^{\mu\nu}$, and we
must therefore consider the gauge transformation
$\delta\Lambda^{\mu\nu}$, which will lead to the gauge field, in more
detail. Using (\ref{eq:brackets1}), (\ref{eq:brackets2}),
(\ref{eq:brackets3}) and (\ref{eq:sph-gauge-tr}), we find
\begin{eqnarray}
  \delta \Lambda^{\mu\nu} =
  \Lambda^\mu\raisebox{-0.35ex}{\scriptsize$\rho$}
  \sigma(\alpha)^{\rho\nu}\label{eq:lorentz.transf}
\end{eqnarray}
where we introduced
\begin{eqnarray}
 \sigma(\alpha)^{\mu\nu}=
  \lambda(\alpha)^{\mu\nu}+
  \tilde{\alpha}(S^{\mu\nu}-\Lambda^0\raisebox{-0.35ex}{\scriptsize$\rho$}
(\Lambda^{0\mu}S^{\nu\rho}-
  \Lambda^{0\nu}S^{\mu\rho}))
\end{eqnarray}
with
\begin{eqnarray}
  \tilde{\alpha}=(\alpha f'(m^2) - \hbox{$\frac{1}{m}$}\alpha_\sigma
  S^{\sigma\lambda}\Lambda^0\raisebox{-0.35ex}{\scriptsize$\lambda$}).
\end{eqnarray}
and where $\lambda(\alpha)^{\mu\nu}$ is given in (\ref{lambdaalpha}).
This transformation formula is more complicated than (\ref{eq:lambdatrans}).
However, for $\Lambda ^{0\nu }$ it reduces to eq. (\ref{eq:lambdatrans}).
Consequently, the determination of  finite gauge transformations of
$\Lambda ^{0\nu }$ reported in sec. IV also applies here.

The equation of motion
\begin{eqnarray}
  \dot{\Lambda}^{\mu\nu}=\{\Lambda^{\mu\nu},H\},\hspace{1.3cm}
  H=\frac{e}{2}\phi+v_\mu\psi^\mu\label{sph.H}
\end{eqnarray}
gives, in analogy with the above calculation
\begin{eqnarray}
\dot{\Lambda}^{\mu\nu}&=&\Lambda^\mu\raisebox{-0.35ex}{\scriptsize$\rho$}
\sigma(v)^{\rho\nu}
  ,\label{eq:lorentz-dot}\end{eqnarray} where $\sigma(v)^{\mu\nu}$ is a
relativistic
angular velocity
defined by
\begin{eqnarray}
  \sigma (v)^{\mu \nu }=\lambda(v)^{\mu\nu}+\tilde{v}(S^{\mu\nu}-
  \Lambda^0\raisebox{-0.35ex}{\scriptsize$\rho$}(\Lambda^{0\mu}S^{\nu\rho}-
  \Lambda^{0\nu}S^{\mu\rho}))\nonumber\\ & &
  \label{sph.gauge.field}
\end{eqnarray}
with $\lambda (v)^{\mu \nu }$ given by (\ref{eq:lambdav}),
and where
\begin{eqnarray}
  \tilde{v}=(e f'(m^2)- \hbox{$\frac{1}{m}$}v_\sigma
S^{\sigma\lambda}\Lambda^0\raisebox{-0.35ex}{\scriptsize$\lambda$}).\label{e
q:sph.v.tilde}
\end{eqnarray}
Eq. (\ref{eq:vtrns}) is found to be valid also in this case, and the Lagrange
multipliers are still constants. We can thus take over
the
analysis of finite gauge transformations of sec. IV.
This means that if we
follow the same way of integrating the transformation equations as in
the point particle case, we can give the same physical interpretation
to the result: The selection of gauge corresponds to a selection of
center of mass.
 We identify $\sigma^{\mu\nu}(v)$
as the gauge field and the Lorentz group $SO(D-1,1)$ as the gauge group.

%%%%%%%%%%%%%%%%%%%%%
\section{Conclusion}
%%%%%%%%%%%%%%%%%%%%%

We have shown that the gauge theory defined by the constraints
(\ref{eq:constraints}) for the relativistic spinning particle is
connected to the problem of selecting a relativistic center of mass.
The gauge theory has the Lorentz group as gauge
group, and the gauge field is the angular velocity
$\lambda^{\mu\nu}(v)$.
The gauge theory of the relativistic
spherical top defined by the constraints (\ref{eq:sph-constraints}) can
be given a similar
physical interpretation; it has the same gauge group
and a similar gauge field as the relativistic spinning particle model.

\vspace{1cm}
{\bf Acknowledgement:} We are grateful to U. J. Quaade
for pointing out the possible relevance of Pryce's papers to the
choice of gauge.

%%%%%%%%%%%%%%%%%%%%%%%%%%%%%%%%%%%%%%%%%%%%%%%%%%%%%%%%%%%%%%%%%

%%%%%%%%%%%%%%%%%%%%%%%%%%%%%%%%%%%%%%%%%%%%%%%%%%%%%%%%%%%%%%%%%

\end{document}